\documentclass[preprint,aps,showpacs]{revtex4}
\usepackage{epsfig}
\usepackage{float}
\RequirePackage{amssymb,amsmath}
\usepackage[latin1]{inputenc}
\usepackage{graphicx}
\usepackage{indentfirst}
\usepackage{color}
\begin{document}

\title{Super-asymmetric fission of heavy nuclei induced by intermediate 
energy protons}

\author{A. Deppman\footnote{Electronic address: adeppman@gmail.com}, 
E. Andrade-II\footnote{Electronic address: esegundo@if.usp.br}, 
V. Guimar\~aes\footnote{Electronic address: valdir@dfn.if.usp.br}, 
G. S. Karapetyan\footnote{Electronic address: ayvgay@ysu.am}}
\affiliation{Instituto de Fisica, Universidade de Sao Paulo,
P. O. Box 66318, 05315-970 S\~ao Paulo, SP, Brazil}

\author{O. A. P. Tavares}
\email{oaptavares@cbpf.br}
\affiliation{Centro Brasileiro de Pesquisas F\'isicas-CBPF/MCTI\\ 
Rua Dr. Xavier Sigaud, 150, 22290-180 Rio de Janeiro - RJ, Brazil}

\author{A. R. Balabekyan}
\email{balabekyan@ysu.am}
\affiliation{Yerevan State University, Faculty of Physics, Alex Manoogian 1, 
Yerevan 0025, Armenia}

\author{N. A. Demekhina}
\email{demekhina@nrmail.jinr.ru}
\affiliation{Yerevan Physics Institute, Alikhanyan Brothers 2, Yerevan 0036, 
Armenia\\
Joint Institute for Nuclear Research (JINR), 
Flerov Laboratory of Nuclear Reactions (LNR), 
Joliot-Curie 6, Dubna 141980, Moscow, Russia}

\author{J. Adam}
\email{iadam@jinr.ru}
\affiliation{Joint Institute for Nuclear Research (JINR), 
Flerov Laboratory of Nuclear Reactions (LNR),
Joliot-Curie 6, Dubna 141980, Moscow, Russia}

\author{F. Garcia}
\email{fermingv@gmail.com}
\affiliation{Departamento de Ci\^encias Exatas e Tecnol\'ogicas-DCET\\ 
Centro de PEsquisas em Ci\^encia e Tecnologias das Radia\c c\~oes - CPqCTR\\
Universidade Estadual de Santa Cruz - UESC, 
 Rodovia Jorge Amado km 16, Ilh\'eus - BA, Brazil}

\author{ K. Katovsky}
\email{k.katovsky@sh.cvut.cz}
\affiliation{Czech Technical University, Department of Nuclear Reactors, 
Prague, Czech Republic}

\begin{abstract}
In this work we present the results for the investigation of intermediate-mass 
fragment (IMF) production with the proton-induced reaction at 660 MeV 
on $^{238}$U and $^{237}$Np target. The data were obtained with the LNR 
Phasotron U-400M Cyclotron at Joint Institute for Nuclear Research (JINR), 
Dubna, Russia.  A total of 93 isotopes, in the mass range of $30 < A < 200$, 
were unambiguously identified with high precision. The fragment production 
cross sections were obtained by means of the induced-activation method 
in an off-line analysis. Mass-yield distributions were derived from the 
data and compared with the results of the simulation code CRISP for 
multimodal fission.  A discussion of the super-asymmetric fragment production 
mechanism is also given.
\end{abstract}

\maketitle

\section{Introduction}
Nuclear dynamics is a complex problem where puzzling aspects of quantum 
mechanics and the natural difficulties of many-body systems are interconnected. 
Besides these factors, the strong interaction, which is not completely 
understood at present time, adds new challenges for calculations in the 
non-perturbative regime. Collective nuclear phenomena, such as fission, 
particle or cluster evaporation, and nuclear fragmentation, offer the 
possibility to  study those complex features of nuclear dynamics. Aside 
from the interest in fundamental nuclear physics, there are many applications 
where the knowledge of fragment formation dynamics in nuclear reactions would 
be helpful. For instance, information on intermediate mass fragment (IMF) 
cross section is relevant for the design of accelerator-driven systems (ADS) 
and radioactive  ion-beam (RIB) facilities, as well as in the study of 
resistance of materials to radiation. 

Fragments in high energy nuclear collisions can be produced by spallation, 
fission, and/or multifragmentation processes. 
J. Hufner \cite{Hufner}, using the 
mass number of the fragments $A$ and their multiplicity $M$ as 
classification parameters,  defined these processes in the following way:

\begin{enumerate}
\item spallation is the process in which only one heavy fragment with a mass 
close to the target mass, $A_{T}$, is  formed (a special case of spallation is 
the so-called deep spallation where $M=1$ but $A \sim \frac{2}{3}A_{T}$);
\item fission is the process in which $M=2$ and $A$ is around $A_{T}/2$;
\item multifragmentation is the process where $M>2$ and $A < 50$.
\end{enumerate}

Emission of light particles, with atomic number $Z \leq 2$, usually dominates 
the yield of reaction products for light target nuclei, while for heavy 
targets spallation and fission residua also give significant contribution. 
Thus, by adopting  the definition that IMF are particles with $A>$ 4 but 
lighter than fission fragments ($A<$100), they can be formed through the 
following processes: i) Fission of nuclei mass number in the range of 
120-130 \cite{Loveland}; ii) Spallation, including the emission of IMF, 
the so-called associated spallation \cite{Yariv};
iii) Multifragmentation of heavy nuclei \cite{Grabez}.

For heavy targets, multifragmentation would be the only mechanism for
 the formation of IMF. Indeed, in Ref. \cite{Grabez}, it was found that in the 
inverse-kinematics reaction of 3.65$A$~MeV~$^{208}$Pb on $^1$H, the formation 
of $^{12}$C nucleus presents the characteristics of multifragmentation, with a 
possible small contribution of binary process. The formation of IMF was 
also observed in measurements at lower energies, see refs. 
\cite{Ricciardi,Kotov,Barz}, but in these cases, the dynamics of the 
process indicates a binary decay with no evidence for multifragmantation. 
Hence, the study of the production of IMF by reactions with heavy target nuclei 
at intermediate energies can give new information on the nuclear dynamics. 
In the present work, our objective is to present new data 
on the measured cross sections for residual nuclei in the IMF region, 
obtained from reactions induced by 660 MeV protons on heavy target nuclei, 
and the corresponding analysis performed with Monte Carlo calculations 
using the CRISP code 
\cite{crisp,Deppman2013}, as described below.

\section{Theoretical aspects of IMF formation}

It is generally assumed that, at intermediate energies, the nuclear reaction 
proceeds in two stages. The first stage would correspond to an incoming
fast projectile colliding with a single nucleon or with several nucleons,
transferring momentum and energy to the nucleus, and leaving the residual
nucleus accompanied by several light particles. The second stage would 
correspond to the decay of the residual nucleus,  which is already in 
statistical equilibrium, by the emission of nucleons or 
clusters of nucleons. At high energies, where the excitation energy per nucleon
of the residual  nucleus is $E_x/A \geq$~3.5~MeV/A, multifragmentation of 
the nucleus can take place. This reaction mechanism differs from evaporation 
since it describes a sudden breakup of the nucleus instead of a successive 
emission of particles.

In the framework described above, the formation of IMF from heavy targets at 
intermediate energies could only be attributed to a process in which fission 
takes place at some point in a long evaporation chain (both pre- and 
post-fission), which is very unlikely. In fact, the fission probability for 
heavy nuclei drops very fast as the mass number decreases 
\cite{Lima,Odilon,Fukahori}, and thus, a long evaporation chain would lead 
to lower fissility nuclei. Another possibility for the formation of IMF would 
be a very long evaporation chain leading to light spallation products. 
This mechanism is limited by the maximum excitation energy allowed for the 
nucleus before multifragmentation becomes dominant, since evaporation 
would cool the nucleus before the IMF region is reached. Increasing the 
excitation energy above the 3.5~MeV/A threshold would only increase the 
contribution from multifragmentation and, in this way, the IMF formed in 
reactions with heavy targets should be dominated by fragmentation products. 
Hence, for excitation energies below the multifragmentation threshold, 
the formation of IMF from heavy nuclei would be very unlikely.

The formation of IMF was observed in the inverse-kinematics reaction 
of  $^{238}$U on proton at 1 AGeV \cite{Ricciardi}, where the cross sections 
for 254 light nuclides in the element range of $7 \leq Z \leq 37$ 
was measured. Based on a detailed study of the experimental kinematic 
information, the authors identified such nuclides as binary decay products 
of a fully equilibrated compound nucleus, whereas clear indications for 
fast breakup processes were absent. Although these result are corroborated 
by those from Refs. \cite{Kotov} and \cite{Barz}, they are in contradiction 
with the scenario described in Ref. \cite{Hufner}.
One possible explanation for the binary production of IMF from reactions 
induced on heavy targets would be by considering highly asymmetric fission   
fragments which can still undergo evaporation to form, at the end, a nuclide 
in the region of IMF. This process would correspond to a modification in the 
classification given by J. Hufner \cite{Hufner} by using a less restrictive 
definition for fission, since in this case, the fragment would have mass 
number quite different from $A_T/2$. This super-asymmetric mechanism would 
corroborate the conjecture that evaporation and fission are manifestation 
of a single mechanism, called binary decay \cite{Moretto,Odilon2}. 
A complete investigation of this possibility involves the description of 
the entire process from the primary interaction of the incident proton up 
to the evaporation of nucleons from the fission fragments. Such a task can 
only be performed through the Monte Carlo method.

Here we used the CRISP code to calculate all of the features of the 
nuclear dynamics during the reaction. CRISP is a Monte Carlo code for 
simulating nuclear reactions \cite{Deppman2004} where it is assumed that 
nuclear reactions can be separated in the two stages already mentioned above: 
the intranuclear cascade and the evaporation/fission process. This code has 
been developed during the last 25 years, and it has been successfully used 
to describe many different reactions. The main characteristic of the 
intranuclear cascade calculation with CRISP is the multi-collisional
 approach \cite{Kodama,Goncalves1997}, where the full nuclear dynamics is 
considered in each step of the cascade. In this process the nucleus is 
modeled by an infinite square-potential which determines the level structure 
for protons and neutrons. The effects of the nuclear potential are present 
in the transmission of the particles through the nuclear surface or through 
an effective mass according to the Walecka mean field approximation 
\cite{Wallecka}. 
The multicollisional calculation is accomplished by constantly updating 
all of the kinematic variables of all particles inside the nucleus, which
opens the possibility for treating more realistically many nuclear phenomena. 
For instance, the anti-symmetrization criteria, which stipulates a strict 
observation of the Pauli Principle, allows the  separation of the 
intranuclear cascade from the thermalization process \cite{Gilson}. 
The effectiveness of such an approach can be verified in 
the processes which are predominantly dependent on the intranuclear cascade 
step, such as kaon production and hypernuclei decay. Such processes have been 
studied with the CRISP code with results compatible with experiments 
\cite{Pina,Israel}. Also, fission of several nuclei has been studied in the 
quasi-deuteron region \cite{Gilson}, where the Pauli blocking mechanism is very 
important in the determination of the residual nucleus.

The evaporation/fission competition was first studied with the CRISP code in 
Refs. \cite{CRISP_PRL,CRISP_NIM,CRISP_CPC}, giving for the first time an 
explanation for the saturation below the unity for the fissility of heavy nuclei
observed in photo-fission experiments at intermediate energies 
\cite{Bianchi,Sanabria}. After, the code was extended to simulate 
reactions at energies up to 3.5~GeV \cite{Gilson2}, showing also good agreement 
with experimental data. After  nuclear thermalization, the competition 
between fission and evaporation processes, which includes neutrons, protons 
and alpha-particles, is determined by the ratios between their respective 
widths according to the Weisskopf model for evaporation and to the 
Bohr-Wheeler model for fission. These ratios are given by:

\begin{align}
 \dfrac{\Gamma_p}{\Gamma_n} = \dfrac{E_p}{E_n} \exp \left\lbrace 2 \left[ 
(a_p E_p)^{1/2} - (a_n E_n)^{1/2} \right] \right\rbrace ,
 \label{gamap}
\end{align}
and
\begin{equation}
 \dfrac{\Gamma_{\alpha}}{\Gamma_n} = \dfrac{2 E_{\alpha}}{E_n} 
\exp \left\lbrace 2 \left[ (a_{\alpha} E_{\alpha})^{1/2} - (a_n E_n)^{1/2} \right] 
\right\rbrace .
 \label{gamaa}  
\end{equation}
for evaporation and by
\begin{align}
 \dfrac{\Gamma_f}{\Gamma_n} = K_f \exp \left\lbrace 2 \left[ (a_f E_f)^{1/2} - (a_n E_n)^{1/2} \right]  \right\rbrace ,
 \label{gamaf}
\end{align}
where,
\begin{align}
 K_f = K_0 a_n \dfrac{\left[ 2(a_f E_f)^{1/2} - 1 \right]}{(4A^{2/3}a_f E_n)} ,
 \label{Kf}
\end{align}
for fission. The parameters $a_i$ stand for the density levels calculated by 
Dostrovsky's parametrization~\cite{Dostrovsky} and $E_i$ is given by
\begin{align}
 \begin{split}
   E_n &= E - B_n, \\
   E_p &= E - B_p - V_p, \\
   E_{\alpha} &= E - B_{\alpha} - V_{\alpha} \\
   E_f &= E - B_f.
 \end{split}
\label{enerEsta}
\end{align}
where $B_n$, $B_p$ and $B_{\alpha}$ are the separation energies for neutrons, 
protons and alphas, respectively, and $B_f$ is the fission barrier. 
$V_i$ stands for the Coulomb potential.

At each $n$th step of the evaporation, the excitation energy of the 
compound nucleus is modified by
\begin{equation}
E^{n}=E^{n-1}-B-V-\varepsilon,
\end{equation}
where $\varepsilon$ is the kinetic energy of the emitted particle.

If the nucleus undergoes fission, the production of fragments is
determined according to the multimodal - random neck rupture
model(MM-NRM) \cite{Brosa1990}, which takes into account the collective 
effects of nuclear deformation during fission by the liquid-drop model and 
single-particle effects by  microscopic shell-model corrections. 
The microscopic corrections create valleys in the space of elongation
and mass number, where each valley corresponds to a different fission 
mode \cite{Brosa1990}.

According to the MM-NRM, the fragment mass distributions are determined by the 
uncorrelated sum of the different fission modes. In principle, it is supposed 
that there are three distinct fission modes for heavy nuclei: symmetric 
Superlong ($S$) mode and two asymmetric modes Standard I ($S1$) and II ($S2$). 
In the Superlong mode, the fissioning system with mass $A_f$ presents itself 
at the saddle-point in an extremely deformed shape with a long neck connecting 
the two forming fragments, which will have masses around $A_{f}/2$. 
The Standard~I mode is characterized by the influence of the spherical 
neutron shell $N_{H}\sim82$ and of the proton shell $Z_{H}\sim50$ in the 
heavy fragments with masses $M_{H}\sim132-134$. The Standard~II mode is 
characterized by the influence of the deformed neutron shell closure $N=86-88$ 
and proton shell $Z_{H}\sim52$ in the heavy fragments with masses 
$M_{H}\sim138-140$.

The fission cross section as a function of mass number is then obtained by 
the sum of three Gaussian functions, corresponding to the three modes 
mentioned above \cite{Younes}:

\begin{align}
 \begin{split}
\sigma (A) = &
\frac{1}{\sqrt{2\pi}}\bigg[\frac{K_{1AS}}{\sigma_{1AS}}
\exp\left({-\frac{(A-A_S-D_{1AS})^2}{2\sigma^2_{1AS}}}\right)+
\frac{K_{1AS}}{\sigma_{1AS}}\exp\left(-\frac{(A-A_S+D_{1AS})^2}
{2\sigma^2_{1AS}}\right)+\\
 & \frac{K_{2AS}}{\sigma_{2AS}}\exp\left({-\frac{(A-A_S-D_{2AS})^2}
{2\sigma^2_{2AS}}}\right)+
\frac{K_{2AS}}{\sigma_{2AS}}\exp\left({-\frac{(A-A_S+D_{2AS})^2}
{2\sigma^2_{2AS}}}\right)+\\
 & \frac{K_S}{\sigma_S}\exp\left({-\frac{(A-A_S)^2}
{2\sigma^2_S}}\right)\bigg],
\label{massTotal}
 \end{split}
\end{align}

\noindent where $A_S$ is the mean mass number determining the center of 
Gaussian functions; and $K_i$, $\sigma_i$, and D$_i$ are the intensity, 
dispersion and position parameters of the $i^{th}$ Gaussian
functions. The indexes $AS$, $S$ designate the asymmetric and symmetric 
components.

The CRISP code works on an event-by-event basis, and therefore, the parameter 
$A_S$ in Eq. (3) is completely determined by the mass of the fissioning 
nucleus $A_f$, that is, $A_S=A_f/2$. 
The positions of the heavy and light peaks of the asymmetric components in 
the mass scale are given by the quantities $A_S$ + D$_{iAS}$ = A$_H$ 
and $A_S$ - D$_{iAS}$ = A$_L$, where $A_H$ and $A_L$ are the masses of the
heavy and light fragment, respectively. 
The values of A$_H$ + A$_L$=2$A_S$ are treated as the mass of the  
undergoing fission nuclei in the respective channel.

One important observable in the  fission process is the charge distribution 
of a given isobaric chain with mass number $A$. It is assumed that this charge 
distribution is well described by a Gaussian function characterized by the 
most probable charge $Z_p$ of an isobaric chain with mass $A$ 
(centroid of the Gaussian function) and the associate width parameter
$\Gamma_z$ of the distribution as following \cite{Kudo,Duijvestijn}:

\begin{eqnarray}
\sigma_{A,Z}=\frac{\sigma_A}{\Gamma_z\pi^{1/2}}
\exp\left({-\frac{(Z-Z_p)^2}{\Gamma_z^2}}\right),
\label{charge}
\end{eqnarray}

\noindent where $\sigma_{A,Z}$ is the independent cross section of the 
nuclide with charge $Z$ and mass $A$.

The values for $Z_p$ and $\Gamma_z$ can be represented as linear 
functions of the mass number of the fission fragments,

\begin{eqnarray}
Z_p=\mu_1+\mu_2A\,,
\label{zp}
\end{eqnarray}
\noindent and
\begin{eqnarray}
\Gamma_z=\gamma_1+\gamma_2 A\,.
\label{width}
\end{eqnarray}

Here $\mu_i$ and $\gamma_i$ were determined by considering a systematic 
analysis of atomic number distributions of fission fragments. The values 
obtained  for all parameters used in the present work are reported in 
Table~\ref{table2}.

It is important to emphasize that  the CRISP code has been used to simulate 
nuclear reactions of several kinds, such as those induced by 
protons~\cite{Evandro_fragdist2,Evandro_fragdist3,crisp_spall_neutr_mult}, 
photons~\cite{Gilson,Gilson2,Gilson3,crisp_ph2002,crisp_ph2009}, 
electrons~\cite{Likhachev2,Likhachev3} or 
hypernuclei~\cite{Israel, Israel2, Krmpotic}, 
with energies from 50~MeV up to 3.5~GeV, and on nuclei with masses going 
from $A=$~12 up to $A=$240 and with several observables: spallation products, 
strange particles, fission products, hyperon-decay particles, fragment mass 
and atomic number distributions. The code has been applied in the study for 
development of nuclear 
reactors~\cite{crisp_ads2,crisp_ads3,crisp_ads5}. 
Thus, the CRISP code is a reliable tool to investigate properties of nuclear 
reactions. 

\section{Experimental Procedure}

In the following we describe how the data present in this work have been 
obtained. A natural uranium target of 0.164 g and 0.0487 mm thick and 
a neptunium target of 0.742 g and 0.193 mm thick were exposed to an accelerated 
proton beam of 660 MeV in energy from the LNR Phasotron, Joint Institute for
Nuclear Research (JINR), Dubna, Russia \cite{Adam}. The proton flux was 
determined by the use of an aluminum monitor with known cross section 
\cite{Cumming}. The monitor, with the same size as the target, was 
irradiated together with the target.
The irradiation time was 27 min and the proton beam intensity was 
about 3 $\times$ 10$^{14}$ protons per min.
 The induced activity of  the targets was measured by two detectors, an HPGe 
detector with  efficiency of 20\% and energy resolution of 1.8 keV 
(1332 keV $^{60}$Co) for the $^{238}$U target and a Ge(Li)  detector with 
efficiency of 4.8\% and energy resolution of 2.6 keV  (1332 keV $^{60}$Co) for 
the $^{237}$Np target. The identification of the reaction products and 
the determination of their production cross section were performed considering
the  half-lives, energies, and intensities  of $\gamma$-transitions of 
the radioactive fragments.

In the absence of a parent isotope, the cross section of fragment production 
for each fragment is determined by using the following equation:

\begin{eqnarray}
\hspace{-0.2cm}\sigma=\frac{\Delta{N}\;\lambda}{N_{p}\,N_{n}\,k\,
\epsilon\,\eta\,(1-\exp{(-\lambda t_{1})})\exp{(-\lambda t_{2})}(1-
\exp{(-\lambda t_{3})})},\label{g1}
\end{eqnarray}

\noindent where $\sigma$ denotes the cross section of the reaction fragment 
production (mb); 
$\Delta{N}$ is the yield under the photo-peak; $N_{p}$ is the projectile 
beam intensity (min$^{-1}$); $N_{n}$ is the number of target nuclei 
(in 1/cm$^{2}$ units); $t_{1}$ is the irradiation time; $t_{2}$ is
the time of exposure between the end of the irradiation and the beginning of 
the measurement; $t_{3}$ is the time measurement; $\lambda$ is the decay 
constant (min$^{-1}$); $\eta$ is the intensity of $\gamma$-transitions; $k$ is 
the total coefficient of $\gamma$-ray absorption in target and detector 
materials, and $\epsilon$ is the $\gamma$-ray-detection efficiency.

When the isotope production in the reaction under investigation is direct 
and independent (I) of the parent nuclei decay, the cross section is 
determined by Eq.~(\ref{g1}). If the yield of a given isotope receives a 
contribution from the $\beta^{\pm}$-decay of neighboring unstable isobar, the
cross section calculation becomes more complicated \cite{Baba}.
If the formation probability of the parent isotope is known from experimental 
data or if it can be estimated on the basis of other sources, then the 
independent cross sections of daughter nuclei can be calculated by the relation:

\begin{widetext}
\begin{eqnarray}
\sigma_{B}=&&\frac{\lambda_{B}}{(1-\exp{(-\lambda_{B}t_{1})})\,
\exp{(-\lambda_{B}t_{2})}(\,1-\exp{(-\lambda_{B} t_{3})})}\times\nonumber\\&&
\hspace*{-1.5cm}
\left.\Biggl[\frac{\Delta{N}}{N_{\gamma}\,N_{n}\,k\,\epsilon\,\eta}-
\sigma_{A}\,f_{AB}\,\frac{\lambda_{A}\,\lambda_{B}}{\lambda_{B}-\lambda_{A}}
\Biggl(\frac{(1-\exp{(-\lambda_{A} t_{1})})\,\exp{(-\lambda_{A} t_{2})}\,(1-
\exp{(-\lambda_{A} t_{3})})}{\lambda^{2}_{A}}\right.\nonumber\\
&&\left.\qquad
-\frac{(1-\exp{(-\lambda_{B} t_{1})})\,\exp{(-\lambda_{B}
t_{2})}\,(1-\exp{(-\lambda_{B} t_{3})})}{\lambda^{2}_{B}}\Biggr)\right.\Biggr],
\end{eqnarray}
\end{widetext}

\noindent where the subscripts $A$ and $B$ in the variables refer to the parent
and daughter nucleus, respectively; the coefficient $f_{AB}$ specifies 
the fraction of nuclei $A$ decaying to nuclei $B$ (this coefficient 
gives the information on how much the $\beta$-decay affects our data; and 
$f_{AB}=1$ when the contribution from the $\beta$-decay corresponds to 100\%); 
and $\Delta{N}$ is the total photo-peak yield associated with the decays of 
the daughter and parent isotopes. The effect of the forerunner can be 
negligible in some limit cases, for example, in the case where the 
half-life of the parent nucleus is very long, or in the case where the 
fraction of its contribution is very small. In the case when parent
and daughter isotopes could not be separated experimentally, the calculated 
cross sections are classified as cumulative ones (C).

\section{Results and Discussion}

The mass distribution, as cross sections as a function of mass number $A$, for 
the fragment produced by 660 MeV proton induced reactions on uranium 
and neptunium targets are shown in Fig. 1 and 2. In both distributions a 
prominent peak is observed around the symmetric fragment mass, which is 
indeed composed  of fragments from the symmetric fission mode.
The distributions, however, present  also a contribution from two asymmetric 
modes \cite{crisp,Deppman2013}. Here we used the code CRISP to interpret
these experimental distributions.

Some previous analysis for the mass distributions of the p+$^{237}$Np and 
p+$^{238}$U systems have been performed for the mass range of 
70$< A <$ 150~\cite{Deppman2013}. In this work, we have added 
new data in the region of intermediate mass fragment (IMF) 
corresponding to $30 < A < 70$ and data in the region of $150<A<200$
from Ref. \cite{Adam}. 
The measured cross sections for the fragments in the mass range 
of $30 < A < 70$ are  listed in Table I, where the quoted errors include 
contributions from those 
associated with the statistical significance of experimental 
results (2-3\%), those in measuring the target thickness (3\%), and those 
in determining the detector efficiency (10\%). 

Usually, studies on the production of fission fragments do not extend to 
light nuclei and the inclusion of this region in our analysis can bring up 
interesting features of the dynamics for fission fragment production.
In fact, theoretical calculations, based on the mass asymmetry parameter 
and fission barrier height~\cite{Ricciardi}, have shown that, for heavy targets 
and for reactions at intermediate or low energies, the cross sections for 
IMF are very  small. As a consequence  most of the experimental observations 
for fission available  in the literature seem to die out for atomic 
numbers below Z=28.

In Figs. 1 and 2 we can observe a shoulder formed in the mass region of
$30 < A < 70$ for both $^{238}$U and $^{237}$Np target distributions. 
The presence of IMF in reactions at energy as low as the one of the present 
study can hardly be attributed to multifragmentation. 
The observation of another shoulder in the region of $170 < A < 200$, for 
both distributions, reinforces the idea of a binary process
 as the origin of the IMF.  These observations, therefore, are in agreement 
with the results obtained by Ricciard {\it et al.}~\cite{Ricciardi}.  
In this work, we present the results of a study performed with the simulation 
code CRISP,  where the new experimental data set in the light mass region is 
described as a possible product of a fission or spallation process. 
To this end, as described in the next section, we have included  an extra 
super-asymmetric fission mode to the code.

%

\section{Super-asymmetric fission mode}

To take into account the possibility of a super-asymmetric fission, we included 
another mode, $S3$, to the CRISP code, which can be
described by the usual Gaussian shape from MM-NRM,

\begin{eqnarray}
\sigma(A)_{3AS}&=&\frac{1}{\sqrt{2\pi}}\left[\frac{K_{3AS}}{\sigma_{3AS}}\exp\left({-\frac{(A-A_{S}-D_{3AS})^{2}}{2\sigma^{2}_{3AS}}}\right)+
\frac{K_{3AS}}{\sigma_{3AS}}\exp\left(-\frac{(A-A_{S}+D_{3AS})^{2}}{2\sigma^{2}_{3AS}}\right)\right],
\label{massAsym}
\end{eqnarray}

As in the case of the three modes previously analyzed in 
Ref.~\cite{Deppman2013}, $K_{3AS}$, $\sigma_{3AS}$ and $D_{3AS}$ are fitting 
parameters which allow us to describe the experimental data for fragments 
produced through the fission channel.
In addition to the fission, we calculated the mass distributions for fragments 
produced by the deep-spallation process.  The results are also shown in 
Figs. 1 and 2, where we observe, that with the inclusion of the 
super-asymmetric mode, the experimental data is well described by the 
fission mechanism according 
to the CRISP calculations. The deep-spallation mechanism gives only a very 
small contribution in the region of heavy fragments, showing that, in fact, the 
super-asymmetric fission is the relevant mechanism for the production of 
fragments in the region of 160$<A<$200.

The best-fit values for the parameters used in the MM-NRM approach are shown in 
Table~II.  The parameters for the $S$, $S1$ and $S2$ modes were already 
discussed in Ref.~\cite{Deppman2013}.  Therefore, we focus here on the 
parameters for $S3$. The super-asymmetric mode contributes with 0.6\% and 
1.2\% of the total fission cross section for the $^{238}$U and for $^{237}$Np 
targets, respectively.  The total fission cross sections are 1140 mb for 
$^{238}$U and 1360 mb for $^{237}$Np. The width for the $S3$ distribution is 
somewhat larger than those  for $S1$ and $S2$, but smaller than that for $S$ 
mode. The most striking feature of the super-asymmetric mode is the mass 
number gap around 60 a.m.u. with respect to the symmetric fragment for both 
cases studied here. Our results confirm that IMF at intermediate energies 
are formed predominantly through a binary process, and that it is described 
by a super-asymmetric fission mode.

As shown in the present work, a good description of the fragment production 
for the full range of mass of 30$<A<$200
was obtained by considering the fission mechanism.
This might indicate that this is, in fact, the actual predominant mechanism.
However,  we can not totally exclude the possibility that a description 
of the experimental data would also be achieved by considering some other sort 
of mechanism, such as evaporation with the inclusion of 
the associated spallation and with emission of fragments heavier than the 
alpha-particle. 

%

\section{Conclusion}

The cross sections for fragments produced by the  proton-induced fission 
on $^{238}$U and $^{237}$Np at 660 MeV were measured at the LNR Phasatron 
(JINR). The fragment mass distributions covering the region of 20$<A<$200, 
allowed the investigation of the production mechanism for the 
intermediate mass fragments (IMF) in the mass range of 20$<A<$70.
It was found that, for each of the IMF observed in the low mass region, 
there was a heavier counterpart in the  region of 170$<A<$200, 
indicating that they are actually produced by a binary process. 
This hypothesis was tested with the use of the CRISP code by including an 
additional super-asymmetric fission mode described according to the MM-NRM 
approach. The results show, indeed, that it is possible to give an accurate 
description of the fragment production in the entire mass region of 20$<A<$200 
by considering the evaporation/fission mechanism in the CRISP code with the 
usual fission modes, namely, one symmetric and two asymmetric, and including 
a fourth super-asymmetric mode. This last mode produces fragments that are 
around 60 a.m.u., far from the symmetric fragment mass, and contributes 
with 0.6\% and 1.2\% to the total fission cross section for $^{238}$U and 
$^{237}$Np, respectively. Our results are in agreement with previous 
results obtained by Ricciardi {\it et al.} ~\cite{Ricciardi} evidencing 
the binary production mechanism for the IMF at intermediate energy nuclear 
reaction.

\section*{Acknowledgment}

G. K. is grateful to the Funda\c c\~ao de Amparo \`a Pesquisa do Estado de
S\~ao Paulo (FAPESP) 2011/00314-0 and 2013/01754-9, and also to 
the International Centre for Theoretical Physics (ICTP) under the Associate 
Grant Scheme. 
A. D. acknowledge the partial support from CNPq under grant 305639/2010-2 
and FAPESP under grant 2010/16641-7.  E. A. acknowledges the support 
from FAPESP under grant 2012/13337-0.
We thank prof. Wayne Seale for reviewing the text.

\medbreak

\newpage
\begin{table}
\caption{Cross section for the measured IMFs products from reaction 
induced by 660 MeV protons on $^{238}$U and $^{237}$Np targets.}
\scalebox{1.0}{
\begin{tabular}{|c|c|c|c|ccc|c|c|c|c|} \hline
Element & Type & \multicolumn{2}{|c|}{Cross section, mb } & & & & 
Element & Type & \multicolumn{2}{|c|}{Cross section, mb }\\ 
\cline{1-4} \cline{8-11}
& & $^{238}$U & $^{237}$Np & & & & & & $^{238}$U & $^{237}$Np \\ 
\cline{1-4} \cline{8-11}
$^{28}$Mg   & C & 0.0043$\pm$4.3E-4 & 0.186$\pm$0.020 &&&& $^{52}$Fe      & I & 6.5E-4$\pm$5.5E-5 & 0.01$\pm$0.01\\ 
\cline{1-4} \cline{8-11}
$^{34m}$Cl  & I & 7.7E-4$\pm$1.5E-5 & 0.08$\pm$0.02   &&&& $^{54}$Mn      & I & 0.11$\pm$0.01     & 0.28$\pm$0.03\\ 
\cline{1-4} \cline{8-11}
$^{38}$S	   & I & 0.007$\pm$1.4E-4  & $\leq$0.08      &&&& $^{55}$Co      & C & 0.02$\pm$0.002    & $\leq$0.036\\
\cline{1-4} \cline{8-11}
$^{38}$Cl   & I & 0.04$\pm$0.008    & $\leq$0.28      &&&& $^{56}$Mn      & C & 0.15$\pm$0.02     & 0.69$\pm$0.07\\
\cline{1-4} \cline{8-11}
$^{39}$Cl   & C & 0.053$\pm$0.005   & 0.023$\pm$0.003 &&&& $^{56}$Co      & I & 0.07$\pm$0.01     & 0.03$\pm$0.006\\
\cline{1-4} \cline{8-11}
$^{41}$Ar   & C & 0.0037$\pm$7.4E-4 & 0.73$\pm$0.07   &&&& $^{56}$Ni      & I & $\leq$0.002       & $\leq$0.007\\
\cline{1-4} \cline{8-11}
$^{42}$K    & C & 0.007$\pm$7.0E-4  & $\leq$0.40      &&&& $^{57}$Co      & I & 0.059$\pm$0.006   & 0.20$\pm$0.02\\
\cline{1-4} \cline{8-11}
$^{43}$K    & C & 0.023$\pm$0.002   & 0.45$\pm$0.06   &&&& $^{57}$Ni      & I & 0.0011$\pm$1.1E-4 & $\leq$0.01\\
\cline{1-4} \cline{8-11}
$^{43}$Sc   & C & 0.012$\pm$0.001   & 0.23$\pm$0.02   &&&& $^{58(m+g)}$Co & I & 0.17$\pm$0.02     & 0.13$\pm$0.02\\
\cline{1-4} \cline{8-11}
$^{44}$Ar   & I & $\leq$2.5E-4      & 0.089$\pm$0.02  &&&& $^{59}$Fe      & C & 0.27$\pm$0.03     & 1.21$\pm$0.12\\
\cline{1-4} \cline{8-11}
$^{44}$K    & I & 0.031$\pm$5.0E-5  & 0.22$\pm$0.04   &&&& $^{60(m+g)}$Co & I & 0.33$\pm$0.03     & 1.70$\pm$0.20\\
\cline{1-4} \cline{8-11}
$^{44g}$Sc  & I & $\leq$0.0025      & $\leq$0.15      &&&& $^{60}$Cu      & C & $\leq$0.006       & $\leq$0.053\\
\cline{1-4} \cline{8-11}
$^{44m}$Sc  & I & 0.065$\pm$0.007   & 0.12$\pm$0.01   &&&& $^{61}$Cu      & C & 0.04$\pm$0.004    & $\leq$0.057\\
\cline{1-4} \cline{8-11}
$^{45}$K    & C &         --        & 0.24$\pm$0.05   &&&& $^{65}$Ni      & I & 0.0017$\pm$1.7E-4 & $\leq$0.04\\
\cline{1-4} \cline{8-11}
$^{46(m+g)}$Sc&I& 0.036$\pm$0.004   & 0.94$\pm$0.09   &&&& $^{65}$Zn      & I & 0.10$\pm$0.01     & 0.87$\pm$0.17\\
\cline{1-4} \cline{8-11}
$^{47}$Ca   & I & 0.024$\pm$0.002   & $\leq$0.067     &&&& $^{65}$Ga     & C & $\leq$0.02        & $\leq$0.043\\
\cline{1-4} \cline{8-11}
$^{47}$Sc   & I & 0.17$\pm$0.02     & 0.63$\pm$0.06   &&&& $^{66}$Ni      & I & 0.015$\pm$0.002   & 0.20$\pm$0.05\\
\cline{1-4} \cline{8-11}
$^{48}$Sc   & I & 0.044$\pm$0.004   & 0.42$\pm$0.04   &&&& $^{66}$Ga      & I & 0.051$\pm$0.005   & $\leq$0.084\\
\cline{1-4} \cline{8-11}
$^{48}$V    & I & 0.022$\pm$0.002   & 0.48$\pm$0.05   &&&& $^{66}$Ge      & I & $\leq$0.003       & $\leq$0.13\\
\cline{1-4} \cline{8-11}
$^{48}$Cr   & I & $\leq$0.0014      & 0.01$\pm$0.001  &&&&  $^{67}$Cu      & C & 0.55$\pm$0.06     & 2.10$\pm$0.21\\
\cline{1-4} \cline{8-11}
$^{49}$Cr   & C & 0.025$\pm$0.005   & 0.073$\pm$0.015 &&&& $^{67}$Ga      & C & 0.06$\pm$0.006    & 0.20$\pm$0.02\\
\cline{1-4} \cline{8-11}
$^{51}$Cr   & C & 0.41$\pm$0.04     & 0.20$\pm$0.02   &&&& $^{69m}$Zn     & I & 0.041$\pm$0.004   & 0.80$\pm$0.16\\
\cline{1-4} \cline{8-11}
$^{52g}$Mn  & C & 0.0015$\pm$1.5E-4 & 0.077$\pm$0.008 &&&& $^{69}$Ge      & C & 0.03$\pm$0.003    & 0.051$\pm$0.012 \\
\cline{1-4} \cline{8-11}
$^{52m}$Mn  & I & 0.0085$\pm$8.5E-4 & 0.205$\pm$0.03  &&&& & & & \\ 
\hline
\end{tabular}}
\label{table1}
\end{table}

\begin{table}
\caption{Parameters for the mass distribution calculations.}
\label{table2}
\scalebox{1.0}{
\begin{tabular}{|c|c|c|} 
\hline
Parameter	& 	$^{238}$U           &	$^{237}$Np\\ 
\hline
$K_{1AS}$	&	(2.0 $\pm$ 5.0)\%  &	(1 $\pm$ 1)\%\\
$\sigma_{1AS}$	&	3.5 $\pm$ 0.8	   &	4.5 $\pm$ 0.4\\
$D_{1AS}$	&	18.5 $\pm$ 0.4	   &	21.3 $\pm$ 0.4\\
$K_{2AS}$	&	(19 $\pm$ 5)\%	   &	(7.7 $\pm$ 0.8)\%\\ 
$\sigma_{2AS}$	&	6.0 $\pm$ 0.5	   &	6.5 $\pm$ 0.6\\
$D_{2AS}$	&	18.0 $\pm$ 0.4	   &	28.3 $\pm$ 0.5\\
$K_{3AS}$	&	(0.5 $\pm$ 0.5)\%  &	(1.2 $\pm$ 0.3)\%\\ 
$\sigma_{3AS}$	&	7.0 $\pm$ 0.5	   &	8.0 $\pm$ 0.7\\
$D_{3AS}$	&	57.0 $\pm$ 0.4	   &	62.0 $\pm$ 0.3\\
$K_S$		&	(56 $\pm$ 5)\%	   &	(79.0 $\pm$ 7.0)\%\\
$\sigma_{S}$	&	13.0 $\pm$ 0.5	   &	13.7 $\pm$ 1.0\\ 
$\mu_{1}$	&	4.1	$\pm$ 0.6  &	5.0 $\pm$ 0.8\\
$\mu_{2}$	&	0.38 $\pm$ 0.01	   &	0.37 $\pm$ 0.01\\
$\gamma_{1}$	&	0.92 $\pm$ 0.08	   &	0.59 $\pm$ 0.02\\
$\gamma_{2}$	&	0.003 $\pm$ 0.001  &	0.005 $\pm$ 0.0002\\
\hline
\end{tabular} }
\end{table}

\newpage

\begin{figure}
\epsfig{file=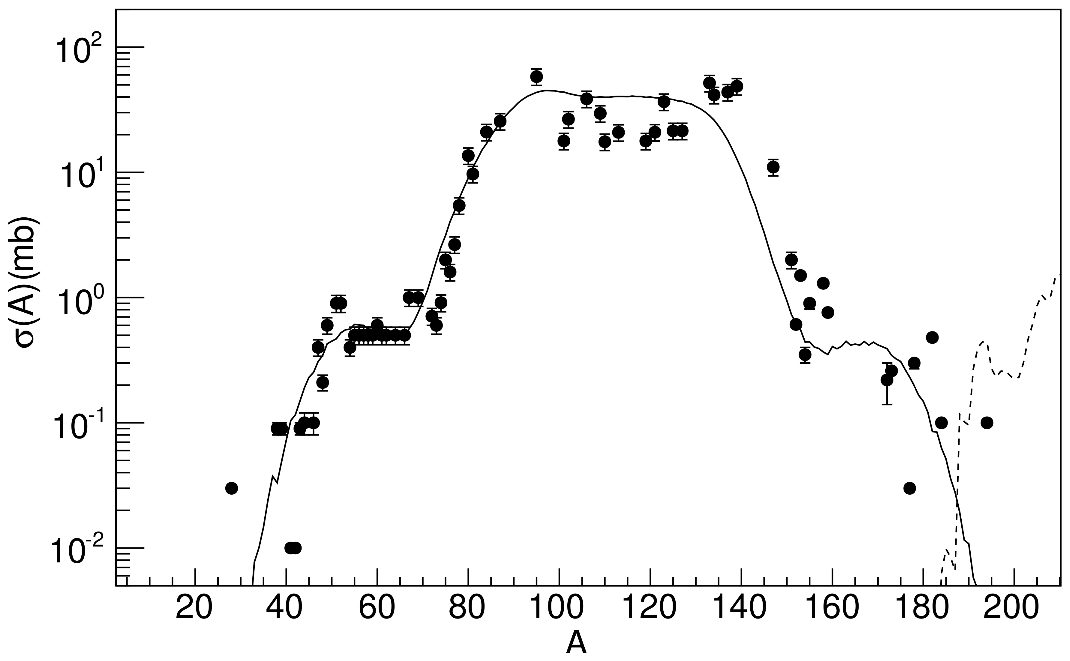,height=9cm,width=12cm,angle=0.}
 \caption{\small Mass distribution of binary-decay products from the 
proton induced reaction at 660 MeV on $^{238}$U target.
Circles represent the measured isobaric cross sections from the 
present work and
from the data taken from Ref. \cite{Adam,Nina3,Nina4}.
The solid line corresponds to the fission process and the dashed line 
represents the results of deep-spallation, both calculated with the CRISP code.}
\end{figure}

\begin{figure}
\epsfig{file=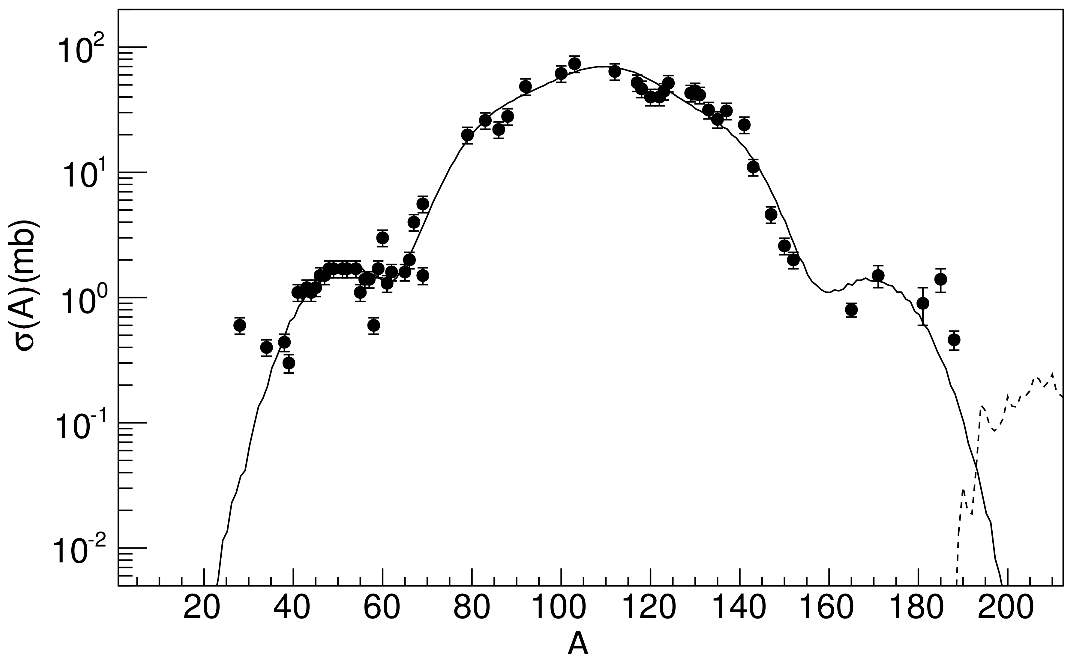,height=9cm,width=12cm,angle=0.}
\caption{\small Idem for the $^{237}$Np target.}
\end{figure}


\bigskip

\begin{thebibliography}{99}

\bibitem{Hufner} J. Hufner, Phys. Rep. $\bf125$, 129 (1985).
\bibitem{Loveland} W. Loveland {\it et al.}, Phys. Rev. C $\bf24$, 464 (1981)
\bibitem{Yariv} Y. Yariv and Z. Fraenkel, Phys. Rev. C $\bf20$, 2227 (1979).
\bibitem{Grabez} B. Grabez, Phys. Rev. C $\bf48$, R2144 (1993).
\bibitem{Ricciardi} M. V. Ricciardi, P. Armbruster, J. Benlliure, {\it et al.}, 
Phys. Rev. C $\bf73$, 014607 (2006).
\bibitem{Kotov} A. A. Kotov, L. N. Andronenko, M. N. Andronenko {\it et al.}, 
Nucl. Phys. A $\bf583$, 575 (1995).
\bibitem{Barz} H. W. Barz, J. P. Bondorf, H. Schulz, {\it et al.}, 
Nucl. Phys. A $\bf460$, 714 (1986).
\bibitem{crisp} A. Deppman, E. Andrade-II, V. Guimaraes, G. S. Karapetyan 
{\it et al.}, Phys. Rev. C {\bf 87}, 054604 (2013).
\bibitem{Deppman2013} A. Deppman, E. Andrade-II, V. Guimaraes, 
G. S. Karapetyan {\it et al.}, Phys. Rev. C {\bf 88}, 024608 (2013).

\bibitem{Lima} D. A. de Lima, J. B. Martins and O. A. P. Tavares, 
Il Nuovo Cim. A $\bf103$, 701 (1990).
\bibitem{Odilon} M. L. Terranova and O. A. P. Tavares. 
Physica Scripta 49, 267 (1994).

\bibitem{Fukahori} T. Fukahori, O. Iwamoto, and S. Chiba, 
{\it Proc. of the 7th Intern. Conf. on Nuclear Criticality Safety, ICNC 2003}, 
JAERI-conf, 2003-019 (pts. 1-2), edited by Nihon
Genshiryoku Kenkyuujo and Nihon Genshiryoku Gakkai (Japan
Atomic Energy Research Institute, Tokai-mura, Japan, 2003), p. 144.
\bibitem{Moretto} L. G. Moretto, Nucl. Phys. A $\bf247$, 211 (1975).
\bibitem{Odilon2} S. B. Duarte, O. A. P. Tavares, F. Guzman 
{\it et al.}, Atom. Data and Nucl. Data Tables 80, 235-299 (2002).
\bibitem{Deppman2004} A. Deppman, S. B. Duarte, G. Silva, {\it et al.}, 
J. Phys. G: Nucl. Part. Phys. ${\bf30}$, 1991-2002 (2004).
\bibitem{Kodama} T. Kodama, S. B. Duarte, K. C. Chung, {\it et al.}, 
Phys. Rev. Lett. $\bf49$, 536 (1982).
\bibitem{Goncalves1997} M. Goncalves, S. dePina, D. A. Lima, {\it et al.}, 
Phys. Lett. B $\bf406$, 1 (1997).
\bibitem{Wallecka} B. D. Serot and J. D. Walecka, 
Adv. Nucl. Phys. $\bf16$, 1 (1986).
\bibitem{Gilson} A. Deppman {\it et al.}, 
J. Phys. G: Nucl. Part. Phys. $\bf30$ 1991 (2004).
\bibitem{Pina} S. de Pina, {\it et al.}, Phys. Lett. B $\bf434$, 1 (1998).
\bibitem{Israel} I. Gonzalez, {\it et al.}, 
J. Phys. G: Nucl. Part. Phys. $\bf38$, 115105 (2011).
\bibitem{CRISP_PRL} A. Deppman {\it et al.}, 
Phys. Rev. Lett. $\bf87$, 182701 (2001).
\bibitem{CRISP_NIM} A. Deppman, {\it et al.}, 
Nucl. Instr. and Meth. in Phys. Res. B $\bf211$, 15 (2003).
\bibitem{CRISP_CPC} A. Deppman,  {\it et al.}, Comp. Phys. Comm. $\bf145$, 
385 (2002).{}
\bibitem{Bianchi} N. Bianchi, {\it et al.}, 
Phys. Lett. B $\bf299$, 219 (1993).
\bibitem{Sanabria} J. C. Sanabria, {\it et al.}, 
Phys. Rev. C $\bf61$, 034604 (2000).

\bibitem{Gilson2} A. Deppman {\it et al.}, 
Phys. Rev. C $\bf73$, 064607 (2006).
\bibitem{Dostrovsky} I. Dostrovsky, P. Rabinowitz and R. Bivins, 
Phys. Rev. 111, 1659 (1958).
\bibitem{Brosa1990} U. Brosa, S. Grossman, and A. Muller, 
Z. Naturforschung $\bf41$, 1341 (1986).
\bibitem{Younes} W. Younes, J. A. Becker, L. A. Bernstein {\it et al.}, 
{\it Nuclear Physics in the 21st Century: International Nuclear Physics 
Conference(INPC 2001)}(AIP, New York, 2001) 
[AIP Conf. Proc. $\bf610$, 673 (2001)].
\bibitem{Kudo} H. Kudo, M. Maruyama, M. Tanikawa {\it et al.}, 
Phys. Rev. C  {\bf 57}, 178 (1998).
\bibitem{Duijvestijn} M. C. Duijvestijn, A. J. Koning {\it et al.}, 
Phys. Rev. C $\bf59$, 776 (1999).
\bibitem{crisp_spall_neutr_mult} S. A. Pereira, {\it et al.}, 
Nucl. Sci. Eng. {\bf 159}. 102 (2008).
\bibitem{Evandro_fragdist2} E. Andrade-II, {\it et al.}, 
J. Phys. G- Nucl. and Part. Phys. {\bf 38} 085104 (2011).
\bibitem{Evandro_fragdist3} E. Andrade-II, J. C. M. Menezes, S. B. Duarte 
{\it et al.}, EPJ Web Conf. {\bf 21} 10001 (2012).
\bibitem{crisp_ph2002} A. Deppman, {\it et al.},  {\it et al.}, 
Phys. Rev. C {\bf 66}, -67601  (2002).
\bibitem{crisp_ph2009} E. Andrade-II, E. Freitas, O. A. P. Tavares 
{\it et al.},  {\it XXXI Workshop on Nuclear Physics in Brazil 
(AIP, New York, 2009)} [AIP Conf. Proc. {\bf 1139}, 64 (2009)].
\bibitem{Gilson3} A. Deppman G. Silva, S. Anefalos,  {\it et al.}, 
Phys. Rev. C {\bf 73}, 064607 (2006).
 \bibitem{Likhachev2} V. P. Likhachev {\it et al.}, 
Nucl. Phys. {\bf 713}, 24 (2003).
 \bibitem{Likhachev3} V. P. Likhachev {\it et al.}, 
Phys. Rev. C {\bf 68}, 014615 (2003).
  \bibitem{Israel2} I. Gonzales, {\it et al.}, 
J. Phys. Conf. Series {\bf 312} 022017 (2011).
 \bibitem{Krmpotic} F. Krmpotic $\textit{et al.}$, 
{\it Nuclear Structure and dynamics 2012} (AIP, New York, 2012)
[AIP Conf. Proc. {\bf 1491}, 117 (2012).
 \bibitem{crisp_ads2} S. Anefalos, {\it et al.}, 
{\it International Conference on Nuclear Data for Science and Technology 2004}
(AIP, New York, 2005) [AIP Conf. Proc. {\bf 769}, 1299 (2005).
 \bibitem{crisp_ads3} S. Anefalos, {\it et al.}, 
Nucl. Sci. and Eng. {\bf 151}, 82 (2005).
 \bibitem{crisp_ads5} A. Deppman, {\it et al.}, 
Sci. and Tech. Nucl. Inst. {\bf 2012}, 480343 (2012).
\bibitem{Adam} J. Adam, K. Katovsky, R. Michel, A. Balabekyan, 
{\it International Conference on Nuclear Data for Science and Technology 2004}
(AIP, New York, 2005) [AIP Conf. Proc. {\bf 769}, 1043 (2005).
\bibitem{Cumming} J. B. Cumming, Ann. Rev. Nucl. Sci. $\bf13$, 261 (1963).
\bibitem{Baba} H. Baba, J. Sanada, H. Araki, {\it et al.}, 
Nucl. Instrum. Methods A $\bf416$, 301 (1998).
\bibitem{Nina3} G. S. Karapetyan, A. R. Balabekyan, N. A. Demekhina, 
and J. Adam, Phys. At. Nucl. $\bf72$, 911 (2009).
\bibitem{Nina4} A. R. Balabekyan, G. S. Karapetyan,  N. A. Demekhina, 
{\it et al.}, Phys. At. Nucl. $\bf73$, 1814 (2010).

\end{thebibliography}
\end{document}